\documentclass[journal]{IEEEtran}
\usepackage[caption=false,font=normalsize,labelfont=sf,textfont=sf]{subfig}
\let\MYorigsubfloat\subfloat
\renewcommand{\subfloat}[2][\relax]{\MYorigsubfloat[]{#2}}
\usepackage{amsmath,epsfig,comment}
\usepackage{amsthm}
\usepackage{tcolorbox}
\usepackage{amssymb}
\usepackage{mathrsfs}
\usepackage{amsmath}
\usepackage{autobreak}
\usepackage{tikz}
\usetikzlibrary{arrows,decorations.markings,automata}
\usepackage{hyperref}
\hypersetup{
	colorlinks=true,
	linkcolor=black,
	filecolor=magenta,
	urlcolor=cyan,
	citecolor=blue
}
\usepackage{cases} 
\usepackage{amssymb,amsmath,cite}
\usepackage{epsfig}
\usepackage{color}
\usepackage{bm}
\usepackage{graphicx}
\usepackage{algorithm}
\usepackage{algorithmic}
\usepackage{multirow}
\usepackage{soul}
\usepackage{extarrows}
\newtheorem{theorem}{Theorem}

\newtheorem{lemma}{Lemma}

\usepackage{float}
\usepackage{booktabs}

\DeclareMathAlphabet\mathbfcal{OMS}{cmsy}{b}{n}

\def\p{\boldsymbol{p}}

\def\y{\boldsymbol{y}}

\usepackage{bm}
\newcommand{\proj}{\mathrm{proj}}
\newcommand{\tr}{\mathrm{tr}}
\newcommand{\prox}{\mathrm{prox}}
\newcommand{\mvec}{\mathrm{vec}}
\newcommand{\bW}{\mathbf{W}}
\newcommand{\bA}{\mathbf{A}}
\newcommand{\bB}{\mathbf{B}}

\newcommand{\bD}{\mathbf{D}}

\newcommand{\bI}{\mathbf{I}}
\newcommand{\bx}{\mathbf{x}}
\newcommand{\bb}{\mathbf{b}}
\newcommand{\by}{\mathbf{y}}
\newcommand{\br}{\mathbf{r}}

\newcommand{\bz}{\mathbf{z}}
\newcommand{\bu}{\mathbf{u}}
\newcommand{\bv}{\mathbf{v}}
\newcommand{\bL}{\mathbf{L}}
\newcommand{\brho}{\bm \rho}
\newcommand{\blam}{\bm \lambda}
\newcommand{\bmu}{\bm \mu}
\newcommand{\Rbb}{\mathbb{R}}
\newcommand{\Cbb}{\mathbb{C}}

\newcommand{\Xcal}{\mathcal{X}}
\newcommand{\Wcal}{\mathcal{W}}
\newcommand{\Ycal}{\mathcal{Y}}

\newcommand{\bQ}{\mathbf{Q}}
\newcommand{\bLam}{\mathbf{\Lambda}}
\newcommand{\bR}{\mathbf{R}}
\newcommand{\bZ}{\mathbf{Z}}
\newcommand{\bT}{\mathbf{T}}
\newcommand{\bH}{\mathbf{H}}
\newcommand{\bS}{\mathbf{S}}
\newcommand{\bU}{\mathbf{U}}
\newcommand{\bh}{\mathbf{h}}

\newcommand{\bp}{\mathbf{p}}

\newcommand{\be}{\begin{equation}}
\newcommand{\ee}{\end{equation}}
\newcommand{\diag}{\mathop{\mathrm{diag}}}
\newcommand{\Diag}{\mathop{\mathrm{Diag}}}
\newcommand{\mat}{\mathrm{mat}}
\newcommand{\Fsf}{\mathsf{F}}
\newcommand{\1}{\mathbf{1}}
\newcommand{\argmin}{\mathop{\mathrm{arg\,min}}}

\newcommand{\iprod}[2]{\left\langle #1, #2 \right\rangle}
\newcommand{\msH}{\dagger}

\newtheoremstyle{noparens}%
{}{}%
{\itshape}{}%
{\bfseries}{.}%
{ }%
{\thmname{#1}\thmnumber{ #2}\mdseries\thmnote{ #3}}
\theoremstyle{noparens}
\allowdisplaybreaks[4]

\begin{document}

\title{A New Adaptive Balanced Augmented Lagrangian Method with Application to ISAC Beamforming Design}

\author{Jiageng Wu, Bo Jiang, Xinxin Li,  Ya-Feng Liu, and Jianhua Yuan
\thanks{J. Wu and X. Li are with the School of Mathematics, Jilin University, Changchun
	130012, China (e-mail: wujg22@mails.jlu.edu.cn; xinxinli@jlu.edu.cn). B. Jiang is with the Ministry of Education Key Laboratory for NSLSCS,
	School of Mathematical Sciences, Nanjing Normal University, Nanjing
	210023, China (e-mail: jiangbo@njnu.edu.cn). Y.-F. Liu is with the State Key
	Laboratory of Scientific and Engineering Computing, Institute of Computational
	Mathematics and Scientific/Engineering Computing, Academy of
	Mathematics and Systems Science, Chinese Academy of Sciences, Beijing
	100190, China (e-mail: yafliu@lsec.cc.ac.cn). J. Yuan is with the School of Sciences, Beijing
	University of Posts and Telecommunications, Beijing 100876, China (e-mail: jianhuayuan@bupt.edu.cn). 
}}

\maketitle

\begin{abstract}
In this paper, we consider a class of convex programming problems with linear equality constraints, which finds broad applications in machine learning and signal processing. We propose a new adaptive balanced augmented Lagrangian (ABAL) method for solving these problems. The proposed ABAL method adaptively selects the stepsize parameter and enjoys a low per-iteration complexity, involving only the computation of a proximal mapping of the objective function and the solution of a linear equation. These features make the proposed method well-suited to large-scale problems. We then custom-apply the ABAL method to solve the ISAC beamforming design problem, which is formulated as a nonlinear semidefinite program in a previous work. This customized application requires careful exploitation of the problem’s special structure such as the property that all of its signal-to-interference-and-noise-ratio (SINR) constraints  hold with equality at the solution and an efficient computation of the proximal mapping of the objective function. Simulation results demonstrate the efficiency of the proposed ABAL method. 	
	
\end{abstract}

\IEEEpeerreviewmaketitle
\section{Introduction}
In this paper, we consider the following linear-equality constrained convex programming problem (in the complex domain):
\begin{equation}\label{eq:gmodel}
\min_{\bu \in \mathbb{C}^n}~f(\bu)\quad \mbox{s.t.} \quad \bD \bu=\bb,  
\end{equation}
where $f:\mathbb{C}^{n}\rightarrow \mathbb{R}\cup\{+\infty\}$ is a proper closed convex function, and $\bD \in\mathbb{C}^{m\times n}$ and $\bb
\in\mathbb{C}^m$ are given complex matrix and vector. Throughout the paper, we assume that the proximal mapping of
the {objective function $f$} in \eqref{eq:gmodel}, which is defined as $$\prox_{\tau f}(\bv):=\argmin_{\bu \in \mathbb{C}^n}\left\{f(\bu)+\frac{1}{2\tau}\|\bu-\bv\|^2\right\},$$ is simple to compute for any given $\bv\in \mathbb{C}^{n}$ and $\tau>0.$ Many problems arising from machine learning and signal processing can be formulated and/or reformulated as problem \eqref{eq:gmodel} \cite{Tomioka2009,aybat2012first,Mota2012,yang2013linearized,Chang2016,liu2022cramer,liu2024survey}.

Various types of algorithms have been proposed for solving problem \eqref{eq:gmodel}; see \cite{douglas1956numerical,hestenes1969multiplier,powell1969method,chambolle2011first,he2021balanced,liu2021acceleration,Ma2023,monteiro2024lowrank} and the references therein. Among all of them, the augmented Lagrangian (AL) method is perhaps the most popular and widely studied one for solving problem \eqref{eq:gmodel}. Specifically, the AL function of problem \eqref{eq:gmodel} is $${\cal{L}}_{\beta}(\bu;\blam)=f(\bu)+\iprod{\bD \bu-\bb}{\blam}+{\frac{\beta}{2}}{\left\|\bD \bu-\bb\right\|}^2,$$  where $\blam\in \mathbb{C}^m$ is the Lagrange multiplier associated with the equality constraint $\bD \bu=\bb$ and $\beta>0$ is the penalty parameter. At the $t$-th iteration, the AL method first (approximately) minimizes    ${\cal{L}}_{\beta}(\bu;\blam^t)$ to obtain the next iterate $\bu^{t+1},$ followed by an update of the multiplier, i.e.,
\begin{equation}\label{alg:ALM}
\begin{cases} 
\bu^{t+1} \in {\argmin \limits_{\bu \in \mathbb{C}^n}{\cal{L}}_{\beta}(\bu;\blam^t)}, \\
\blam^{t+1}=\blam^t+\beta\left(\bD\bu^{t+1}-\bb\right).
\end{cases}
\end{equation} 
In general, the efficiency of the AL method heavily depends on that of solving the AL subproblem (i.e., the $\bu$-subproblem) to get the next iterate. The efficiency of the AL method also relies on the choice of the penalty parameter. The recent progress on reducing the computational cost of solving the AL subproblem \cite{he2021balanced,Ma2023,LiuLMOR} and adaptively selecting the penalty parameter \cite{Lorenz2019,Zhang2024}  in order to improve the efficiency of the AL method will be briefly reviewed in the next section. Combining these existing techniques and taking a step further, we propose \emph{a new adaptive balanced AL (ABAL) method} for solving problem \eqref{eq:gmodel}, which is the first contribution of this paper. The proposed ABAL method not only adaptively selects the stepsize parameter but also requires only computing a proximal mapping of the objective function and solving a linear equation at each iteration, making it well suitable to solve large-scale ill-conditioned problems.

Integrated sensing and communications (ISAC) has been recognized as one of six usage scenarios of 6G in the Recommendation \cite{ITU-R2023} for IMT-2030. Unlike traditional independent communication and radar systems, ISAC aims to achieve dual communication and sensing functionalities within a unified system via sharing wireless resources (e.g., time, frequency, power, and beam) \cite{liu2022integrated}. To achieve the above goal, various resource allocation and in particular beamforming (also called precoding) design problems have been formulated as optimization problems in the literature (e.g., \cite{liu2018toward,vijay2019toward,liu2020joint,liu2022cramer,liu2024random,liu2024survey,wu2024efficient,zou2024energy}). As the second contribution of this paper, we \emph{custom-apply the proposed ABAL method to solve the ISAC beamforming design problem} \cite[Problem (36)]{liu2022cramer}. To facilitate the application of the ABAL method, we first show that all signal-to-interference-and-noise-ratio (SINR) constraints of the beamforming design problem are satisfied with equality at the solution. Compared against the way of solving the problem by calling SeDuMi as in \cite{liu2022cramer}, the customized application of our proposed ABAL method is  2.8 to 600 times faster.

\textit{Notation}. 
The inner product  between matrices $\bA$ and $\bB$ of the same size is given by $\iprod{\bA}{\bB}=\tr(\bA^\msH\bB)$, where $\tr(\cdot)$ denotes the matrix trace and $\bA^\msH$ is the conjugate transpose of $\bA$. The above inner product operator also applies to vectors. Let $\mathbf{0}$, $\mathbf{1}$, and $\bI$ denote the all-zero, all-one, and identity matrices of appropriate sizes.  Let $\mathbb{H}^{N \times N}$ and $\mathbb{H}_+^{N \times N}$ denote the sets of all Hermitian matrices and Hermitian positive semidefinite matrices of size $N \times N$, respectively. For a closed convex set $\mathcal{X}$, $\proj_\mathcal{X}(\cdot)$ is the projection operator onto $\mathcal{X}$ and $\mathbb{I}_{\mathcal{X}}(\cdot)$ is the indicator function of $\mathcal{X}$.  For a proper closed convex function $f$, its subdifferential is denoted by $\partial f.$  
\section{A New Adaptive Balanced Augmented Lagrangian Method}In this section, we first briefly review the balanced AL (BAL) method developed in \cite{he2021balanced} for solving problem \eqref{eq:gmodel} in Section \ref{subsection:bal}. Then, in Section \ref{subsection:abal}, we propose a new ABAL method for solving problem \eqref{eq:gmodel}, which equips the BAL method with an adaptive stepsize.
\subsection{BAL Method and Our Motivation}\label{subsection:bal}

The AL method, which dates back to \cite{hestenes1969multiplier,powell1969method}, is one of the most efficient approaches to solving problem \eqref{eq:gmodel}. However, as shown in \eqref{alg:ALM}, the $\bu$-subproblem in the AL method generally does not admit a closed-form solution due to the presence of both the objective function $f$ and the coefficient matrix $\bD$, and often needs to be solved by an iterative algorithm, which is computationally expensive. To address this issue, the work \cite{he2021balanced} proposed a BAL method, which separates the two difficulties by ``moving'' the terms related to $\bD$ into the $\blam$-subproblem. The BAL method is presented as follows:
\begin{equation}\label{alg:BALM-p}
\begin{cases}
\mathbf{u}^{t+1} =\prox_{\tau f}(\bu^t-\tau\bD^\msH\blam^t),\\
{\mathbf{p}^{t+1}} =\mathbf{D}\left(2 \mathbf{u}^{t+1}-\mathbf{u}^t\right)-\mathbf{b}, \\
\boldsymbol{\lambda}^{t+1} =\boldsymbol{\lambda}^t+(\gamma \tau)^{-1}\left(\mathbf{D} \mathbf{D}^{\msH}+ \theta^2\mathbf{I}\right)^{-1}{\mathbf{p}^{t+1}}.
\end{cases} 
\end{equation}
In the above, 
$\tau>0$ is the primal stepsize, $\gamma$ is the dual stepsize, and $\theta>0$ is a preset small parameter. The parameter $\gamma$ is set to be 1 in \cite{he2021balanced}, and is extended to any number in $[3/4, +\infty)$ in \cite{Ma2023}. The worst-case $\mathcal{O}(1/t)$ convergence rate and global convergence of the BAL method are established in \cite{he2021balanced,Ma2023}.

Two remarks on the BAL method in \eqref{alg:BALM-p} are in order. First, as can be seen from \eqref{alg:BALM-p}, both of the $\bu$-subproblem and the $\blam$-subproblem in the BAL method  \eqref{alg:BALM-p} are simple. In particular, the $\bu$-subproblem in \eqref{alg:BALM-p} admits a closed-form solution, which is in sharp constrast to the $\bu$-subproblem in the classical AL method \eqref{alg:ALM}; solving the $\blam$-subproblem in \eqref{alg:BALM-p} is equivalent to solving the following linear equation
\begin{equation}\label{dualupdate}
\left(\bD\bD^\msH +\theta^2 \mathbf{I}\right)\left(\blam-\blam^t\right)= (\gamma\tau)^{-1}\mathbf{p}^{t+1},\end{equation} which is efficiently solvable. More importantly, in many cases of interest (including the case to be presented in Section \ref{sec:beamforming}), the coefficient matrix $\mathbf{D}$ exhibits a special structure, which further enables more efficient solution of the  $\blam$-subproblem.

Second, the high computational efficiency of the BAL method \eqref{alg:BALM-p} depends on a good choice of the stepsize parameter $\tau$, which is typically chosen manually in literature (e.g., \cite{Ma2023}). 
However, manual tuning of the stepsize parameter $\tau$ is computationally intensive and lacks robustness, highlighting the need for an adaptive strategy to select $\tau$.  Recently, the work \cite{Zhang2024} proposed an accelerated BAL method with a dynamic stepsize $\tau$, but their approach relies on the strong convexity assumption of $f$, which does not apply to our interested problem. 

Given this background, we are motivated to develop a new BAL method with an adaptive choice of the stepsize parameter $\tau$ for solving large-scale problem \eqref{eq:gmodel} without the strong convexity assumption.

\subsection{Proposed ABAL Method}\label{subsection:abal}
In this subsection, we propose a new ABAL method for solving problem \eqref{eq:gmodel}. This is achieved based on the fact that the BAL method can be interpreted as a DRS method \cite{douglas1956numerical} and the application of the non-stationary DRS framework \cite{Lorenz2019} to the BAL method. 

More specifically, using the lift technique introduced in \cite{o2020equivalence}, we can equivalently rewrite problem \eqref{eq:gmodel} as 
\begin{equation} \label{prob:lift1}
\min_{\bu \in \mathbb{C}^n}~f(\bu)\quad \mbox{s.t.} \quad \bD \bu + \theta \bv =\bb, \bv = \mathbf{0}.  
\end{equation}
Denote $h(\bu,\bv) = f(\bu) + \mathbb{I}_{\{\mathbf{0}\}}(\bv)$ and $g(\bu,\bv) = \mathbb{I}_{\{\bb\}}(\bD \bu + \theta \bv).$  
Then we can further rewrite problem \eqref{prob:lift1} as 
\begin{equation}\label{eq:gmodel:lift}
\min_{\bu \in \mathbb{C}^n, \bv \in \mathbb{C}^m}\, h(\bu, \bv) + g(\bu,\bv).
\end{equation}
By the optimality condition, solving problem \eqref{eq:gmodel:lift} is equivalent to finding  a zero of the sum of two maximal monotone operators, i.e.,  
\begin{equation}\label{prob:maxAB}
\mathbf{0} \in (\partial h + \partial g)(\bu,\bv).
\end{equation}
Now we can apply the non-stationary DRS method \cite{Lorenz2019} to solve problem \eqref{prob:maxAB}, which yileds the ABAL method, as outlined in Algorithm \ref{algo:abalmp}.  
The detailed derivation is similar to that of tuning-free PDHG in \cite{wang2024tuning} and thus is omitted. The global convergence of Algorithm \ref{algo:abalmp} follows from \cite[Theorem 3.2]{Lorenz2019}. Note that the initial value $\tau_0$ is set to be $1$ in \cite[Theorem 3.2]{Lorenz2019}. Our result allows $\tau_0$ to be chosen as any positive number.

\begin{theorem}\label{thm:NDRS_conv}
Let $\{\omega_t\} \subset (0,1]$ be a sequence such that  $\sum_{t = 0}^{+\infty} \omega_t < +\infty$ and $\omega_0 = 1$. Then the sequence $\{\bu^t\}$ generated by Algorithm \ref{algo:abalmp} converges to an optimal solution $\bu^*$ of problem \eqref{eq:gmodel}. 
\end{theorem}

Before leaving this section, we make some remarks on the comparison between our work and the one in \cite{wang2024tuning}. First, the work \cite{wang2024tuning} applied the non-stationary DRS method to solve a different lift problem from problem \eqref{prob:lift1}, where the term $\theta \bv$ in \eqref{prob:lift1} is replaced by $\mathbf{T} \bv$ with $\mathbf{T}$ being a carefully chosen coefficient matrix. Second, the key difference between the tuning-free PDHG method in \cite{wang2024tuning} and our proposed ABAL method lies in their update of the dual variable. In particular, the dual variable in \cite{wang2024tuning} is updated as follows: \[
\blam^{t + 1} = \blam^t + (\upsilon \tau_t \|\bD\|_2^2)^{-1} \p^{t+1}\quad {\mbox{with}~~ \upsilon \geq 1}.
\]
The above update of the dual variable can be interpreted as an approximation of our update rule in {Algorithm \ref{algo:abalmp}} with the matrix $\bD\bD^\msH + \theta^2 \bI$ approximated by {$\upsilon\|\bD\|_2^2 \bI$.} This approximation simplifies the dual update but also significantly slows down the convergence of the corresponding algorithm, as can be seen clearly in Section IV. 

\begin{algorithm}[!tbp]
	\caption{Proposed ABAL method for solving problem \eqref{eq:gmodel}}
	\label{algo:abalmp}
	\begin{algorithmic}
		\STATE \textbf{Input:} $\bu^0,\blam^0, \{\omega_t\}$, $\tau_0 > 0$, $\theta > 0$, $0 < \underline{\eta} < \overline{\eta}$.
		\FOR{$t=0,1,\ldots$}
        \STATE $\widetilde \bu^{t} = \bu^t-\tau_{t-1}\bD^\msH\blam^t$, ~$\bu^{t+1}=\prox_{\tau_{t-1}}(\widetilde \bu^{t})$,\\[3pt]
		\STATE
	$\eta_t=\proj_{\left[\underline{\eta},\overline{\eta}\right]}\left({\| 
			\bu^{t+1}\|}\bigg/{\left\|\begin{pmatrix}\bu^{t+1}-\widetilde \bu^{t}\\ \theta \tau_{t-1}  \blam^t \end{pmatrix}\right\|}
   \right)$,\\[3pt]
		\STATE
		$\kappa_t = 1 - \omega_t + \omega_t \eta_t$,~$\tau_t = \kappa_t \tau_{t-1}$,\\[3pt]
		\STATE
		$\mathbf{p}^{t+1} 
		=\bD\left(\bu^{t+1} + \kappa_t (\bu^{t+1}- \bu^t)\right)-\bb$,\\[2pt]
		\STATE
		$\blam^{t+1}=\blam^t+ \tau_t^{-1} \left(\bD\bD^\msH+\theta^2 \bI\right)^{-1}\mathbf{p}^{t+1}$. 
		\ENDFOR
	\end{algorithmic}
\end{algorithm}

\section{ISAC Beamforming Design}
\label{sec:beamforming}
In this section, we apply the proposed ABAL method to solve the ISAC beamforming design problem \cite[Problem (36)]{liu2022cramer}. Before doing it, we first introduce the beamforming design problem and analyze its special structures to facilitate the application of the ABAL method.

\subsection{Problem Reformulation and Structure}
Consider the same scenario as in \cite{liu2022cramer}, where a base station (BS) equipped with $N>1$ antennas serves $K$ single-antenna users and at the same time detects an (extended) target.  For \(k = 1,2, \ldots, K\), let $\bh_k \in \mathbb{C}^N$ denote the channel between the BS and the $k$-th user and $\Gamma_k$ denote the SINR target required by the $k$-th user. Let $\sigma_C^2$ be the noise level and $P_T$ be the power budget of the BS. Then the ISAC beamforming design problem \cite[Problem (36)]{liu2022cramer} of minimizing the Cram\'er-Rao bound subject to all users' SINR constraints and the BS's power budget constraint is given as follows:
\begin{equation}\label{eq:prob}
\begin{aligned}
\min_{\left\{\bW_k\right\}_{k=1}^{K+1}}~  & \tr\left(\left(\sum\nolimits_{k=1}^{K+1} \bW_k\right)^{-1}\right) \\
\displaystyle \mathrm{s.t.}~\quad &
\rho_k \iprod{\bQ_k}{\bW_k}- \sum\nolimits_{i=1}^{K+1} \iprod{\bQ_k}{\bW_i} \geq \sigma_C^2,\, \forall\, k, \\
& \displaystyle\sum\nolimits_{k=1}^{K+1} \tr\left(\bW_k\right) \leq P_T,\bW_k \in \mathbb{H}_+^{N \times N},\,\forall\,k,
\end{aligned}
\end{equation}
where $\bQ_k = \bh_k\bh_k^\msH, \rho_k = 1 + \Gamma_k^{-1},$ and $\bW_k \in \mathbb{H}_+^{N \times N}$ is the relaxed matrix variable from the beamforming vector to be designed for the $k$-th user for $k=1,2,\ldots,K.$ It has been shown in \cite{liu2020joint,liu2022cramer} that problem \eqref{eq:prob} must have a rank-one solution of $\left\{\bW_k\right\}_{k=1}^K.$ In this paper, we focus on solving problem \eqref{eq:prob}.

Problem \eqref{eq:prob} is a nonlinear semidefinite program (SDP) with $K+1$ inequality constraints, which are generally more challenging to handle than equality constraints. Fortunately, by leveraging the special structure of problem \eqref{eq:prob}, we can show, in the following Theorem \ref{thm:conv}, that there exists a solution of problem \eqref{eq:prob} at which all (SINR and power budget) inequality constraints hold with equality. This property will significantly faciliate the application of the proposed ABAL method to solve problem \eqref{eq:prob}. The proof of Theorem \ref{thm:conv} can be found in Appendix  \ref{section:equiv}.

\begin{theorem}\label{thm:conv}
There exists a solution to problem \eqref{eq:prob} at which all inequality constraints hold with equality. 
\end{theorem}

It follows from Theorem \ref{thm:conv} that problem \eqref{eq:prob} is equivalent to the following problem: 
\begin{equation}\label{eq:prob_eq}
\begin{aligned}
\min_{\bW}~  &\tr \left(\left(\sum\nolimits_{k=1}^{K+1} \bW_k\right)^{-1}\right)+\mathbb{I}_ \mathcal{\mathcal{W}}(\bW)\\
\mathrm{s.t.}~~ & 
\rho_k \iprod{\bQ_k}{\bW_k}- \sum\nolimits_{i=1}^{K+1} \iprod{\bQ_k}{\bW_i} = \sigma_C^2,\,\forall\,k, \\
\end{aligned}
\end{equation} where $\bW = \begin{bmatrix} \bW_1, \bW_2, \ldots, \bW_{K+1} \end{bmatrix}^\msH$ and ${\cal W} = \big\{\bW\in \Cbb^{(K+1)N \times N} \mid  \sum\nolimits_{k = 1}^{K+1} \tr(\bW_k)= P_T,\bW_k \in \mathbb{H}_+^{N \times N}, k = 1, 2, \ldots, K + 1\big\}$.

\subsection{Customized Application}\label{section:customized}
In this subsection, we apply the proposed ABAL method (i.e., Algorithm \ref{algo:abalmp}) to solve problem \eqref{eq:prob_eq}. To achieve an efficient application, it is expected that each subproblem therein is computationally efficient. However, for problem \eqref{eq:prob_eq}, solving the $\mathbf{W}$-subproblem is computationally expensive, as it requires computing the proximal mapping of the objective function of problem \eqref{eq:prob_eq}, which is the sum of the trace of the inverse of a matrix and the indicator function of a convex set. To mitigate this difficulty, we introduce an auxiliary variable $\mathbf{Z} = \sum_{k = 1}^{K+1} \bW_k$ into problem \eqref{eq:prob_eq} and further reformulate it as the following compact form that is more favorable for algorithmic design:
\begin{equation}\label{eq:prob2}
\begin{aligned}
\min_{\bW,\,\bZ}\quad& h(\bZ):= \tr \left(\bZ^{-1}\right)+\mathbb{I}_ \mathcal{\mathcal{W}}(\bW)  \\
\mathrm{s.t.}\quad & \mathcal{A}(\bW, \bZ) = \sigma_C^2 \mathbf{1}, ~ \mathcal{B}(\bW,\bZ) = \mathbf{0},
\end{aligned}
\end{equation} where two linear operators $\mathcal{A}$ and $\mathcal{B}$ are defined as follows: $\mathcal{A}: \mathbb{C}^{(K+1)N \times N} \times \mathbb{C}^{N \times N} \to \mathbb{R}^K$  
with $(\mathcal{A}(\bW, \bZ))_k = \rho_k \iprod{\bQ_k}{\bW_k} - \iprod{\bQ_k}{\bZ}$ for $k = 1, 2, \ldots, {K}$ and $\mathcal{B}: \mathbb{C}^{(K+1)N \times N} \times \mathbb{C}^{N \times N} \to \mathbb{R}^{N \times N}$ with $\mathcal{B}(\bW, \bZ) = \sum\nolimits_{k=1}^{K+1} \bW_k - \bZ$. 

By vectorizing the variables $\bW$ and $\bZ$ and choosing an appropriate $f$, it is clear to see that problem \eqref{eq:prob2} is an instance of problem \eqref{eq:gmodel} with $n = (K+2)N^2$ and $m = K + N^2$. Consequently, we can apply 
Algorithm \ref{algo:abalmp} to solve \eqref{eq:prob2}. As will be seen in Section \ref{section:implementation}, the complexity of computing the proximal mapping of the corresponding $f$ is $\mathcal{O}((K + 2)N^3)$. Moreover, at first glance, updating the variable $\blam$ seems to require solving a linear system of size $(K + N^2) \times (K + N^2)$, which generally incurs  a complexity of $\mathcal{O}((K + N^2)^3)$. Fortunately, by fully leveraging the special structure of the coefficient matrices associated with $\mathcal{A}$ and $\mathcal{B}$, we can significantly reduce this complexity to $\mathcal{O}(K^3 + KN^2)$. 

 The $t$-th iteration of the customized ABAL algorithm for solving problem  \eqref{eq:prob_eq} is presented below, with detailed derivations provided in Appendix \ref{appendix:customized}. 
\begin{equation}
\label{equ:customized}
\begin{cases}
\widetilde \bW^t = \bW^t-\tau_{t-1}\Diag([\brho\odot \bmu^t;0])\bQ
   -\tau_{t-1}\mathbf{1}\otimes \bLam^t,  \\[2pt]
\bW^{t+1}=\proj_{\Wcal}\big(\widetilde \bW^t\big),\quad \Delta \bW^t=\bW^{t+1}- \widetilde \bW^t,\\[2pt]
\widetilde{\bZ}^t = \bZ^t + \tau_{t-1} \bH^\msH \Diag(\bmu^t) \bH  + \tau_{t-1} \bLam^t, \\[2pt]
\bZ^{t+1}=\prox_{\tau_{t-1}h}\big(\widetilde{\bZ}^t\big),\quad \Delta \mathbf{Z}^t=\bZ^{t+1}- \widetilde \bZ^t,\\[2pt]
\eta_t = \proj_{[\underline{\eta}, \overline{\eta}]}\left({\frac{\sqrt{\|\bW^{t+1}\|_\Fsf^2 + \|\bZ^{t+1}\|_\Fsf^2}}{\sqrt{\|\Delta \bW^t\|_\Fsf^2+ \|\Delta \bZ^t\|_\Fsf^2+\theta^2\tau_{t-1}^2(\|\bLam^t\|_\Fsf^2+\|\bmu^t\|^2)}}}\right), \\[2pt]
\kappa_t = 1-\omega_t+\omega_t \eta_t,~ \tau_t = \kappa_t \tau_{t-1},\\[2pt]
\br^{t+1} =  (1 + \kappa_t) \mathcal{A}(\bW^{t+1}, \bZ^{t+1}) - \kappa_t \mathcal{A}(\bW^{t}, \bZ^{t}) - \sigma_C^2 \mathbf{1},\\[2pt]
\bR^{t+1} ={}  \left(1 + \kappa_t \right)\mathcal{B}(\bW^{t+1}, \bZ^{t+1}) - \kappa_t \mathcal{B}(\bW^{t}, \bZ^t),\\[2pt]
\bmu^{t+1}={} \bmu^t +     \frac{\bL^{-1}\left(\tau_t\br^{t+1} - \diag(\bH^\msH \bR^{t+1} \bH) \odot (\brho + \1)\right)}{\tau_t(K + 2 + \theta^2)}, \\[2pt]
\bLam^{t+1} ={}\bLam^t +  \frac{\bR^{t+1} - \tau_t \bH^\msH \Diag( (\bmu^{t +1 } - \bmu^t) \odot (\brho + \mathbf{1})) \bH}{\tau_t(K + 2 + \theta^2)}.
\end{cases}
\end{equation} In \eqref{equ:customized}, $\bmu \in \Rbb^K$ and $\bLam \in \mathbb{H}^{N \times N}$ are the corresponding Lagrange multiplier matrices associated with the linear equality constraints in problem \eqref{eq:prob2}; $\bQ = \begin{bmatrix} \bQ_1, \bQ_2, \ldots,\bQ_K \end{bmatrix}^\msH$; $\bL= \theta^2 \bI +  \left|\bH^\msH \bH \right|^2\odot (\Diag(\brho \odot \brho) + \bS) \in \mathbb{R}^{K \times K}$,
where $\bH = [\bh_1,\bh_2,\ldots,\bh_K]$ and the $(i,j)$-th entry of $\bS$ is given by 
$s_{ij} =  1 -{(\rho_i + 1)(\rho_j + 1)}/{(K + 2 + \theta^2)}$. In the above, {$\otimes$ and} $\odot$ {represent the Kroneker and} Hadamard product operators, respectively, and $|\,\cdot\,|^2$ is applied elementwise.

The exploitation of the special structures of problem \eqref{eq:prob} plays a central role in applying the proposed ABAL algorithm for efficiently solving the problem. The structures of problem \eqref{eq:prob} exploited here include the solution structure in Theorem \ref{thm:conv}, the tracable proximal mapping of the trace-inverse function, and the block structure of the matrix $\bD.$ These structures can also be utilized in applying the BAL method  \cite{he2021balanced} to solve problem \eqref{eq:prob}, as done in Section IV.

\subsection{Implementation Details} \label{section:implementation} The dominant computational cost in \eqref{equ:customized}  lies in computing $\bW^{t+1}$ and $\bZ^{t+1}$, which involves performing $K+2$ eigenvalue decompositions of $N \times N$ Hermitian matrices, as shown in Lemmas \ref{thm:Wsub} and \ref{thm:Zsub}. The total complexity is $\mathcal{O}((K+2)N^3)$. 
\begin{lemma}\label{thm:Wsub}
Let $\widetilde \bW_k = \bU_k \bm\Sigma_k \bU_k^\msH$ be the eigenvalue decomposition of $\widetilde \bW_k$ for $k = 1,2, \ldots, K+1$, where $\bU_k \in \mathbb{C}^{N \times N}$ is a unitary matrix and $\bm\Sigma_k$ is diagonal with real entries.  Then, {for $k = 1, 2, \ldots, K + 1$}, we have 
	\begin{equation}
	\bW_k^{t+1} =\bU_k \mathrm{Diag}\left( P_T \cdot \proj_{\Delta}\left(\frac{\diag(\bm\Sigma_k)}{P_T}\right)\right) \bU_k^\msH. 
	\end{equation}
Here, $\Delta$ is the standard simplex in $\mathbb{R}^N$.
\end{lemma}

\begin{lemma}\label{thm:Zsub}
Let $\widetilde{\bZ}^t = \bU \bm{\Sigma} \bU^\msH$ be the eigenvalue decomposition of $\widetilde{\bZ}^t$, where $\bU_k \in \mathbb{C}^{N \times N}$ is a unitary matrix and $\bm\Sigma$ is diagonal with real entries. Then, $\bZ^{t + 1} = \bU \Diag(\bz^*) \bU^\msH$, where $\bz^* \in \mathbb{R}^N$ with its $i$-th element $z_i^*$ being the largest positive root of the cubic polynomial $x^3 - \Sigma_{ii} x^2 - \tau_{t-1} = 0$ for $i = 1,2, \ldots, N$.
\end{lemma}

\section{Numerical Results}
\begin{table}[t]
 	\label{tab:comp:newx}
 	\centering
 	\caption{Comparison results of \textnormal{SeDuMi}, \textnormal{TF-PDHG}, \textnormal{BAL-C}, and \textnormal{ABAL}, denoted as ``\textnormal{a}'', ``\textnormal{b}'', ``\textnormal{c}'', and ``\textnormal{d}'' for brevity.}
 \setlength{\tabcolsep}{3pt}	
  \begin{tabular}{ccccccccccccccc}
 		\toprule
 		 &  \multicolumn{4}{c}{f-gap}  && \multicolumn{4}{c}{iter}  &&  \multicolumn{4}{c}{time} \\
    \cline{2-5}   \cline{7-10}  \cline{12-15}
  $K$ & a & b & c &d && a & b & c &d && a & b & c &d  \\[-2pt]
  \midrule
  \multicolumn{15}{c}{$N = 32$} \\ 
4 & 5e-08 & -- & 1e-07 & 0e+00 && 19 & -- & 708 & 558 && 25.3 & -- & 1.8 & 1.4 \\ 
6 & 2e-07 & -- & 3e-07 & 0e+00 && 22 & -- & 961 & 807 && 29.2 & -- & 3.1 & 2.6 \\ 
8 & 3e-07 & -- & 0e+00 & 4e-07 && 22 & -- & 1449 & 1074 && 31.1 & -- & 5.7 & 4.1 \\ 
10 & 6e-07 & -- & 7e-07 & 0e+00 && 21 & -- & 2269 & 1679 && 30.3 & -- & 10.9 & 6.7 \\ 
12 & 0e+00 & -- & 5e-07 & 4e-07 && 22 & -- & 3075 & 2201 && 32.2 & -- & 16.9 & 11.4   
\\[-1pt]
\midrule 
  \multicolumn{15}{c}{$N = 64$} \\
4 & -- & -- & 4e-09 & 0e+00 && -- & -- & 1602 & 535 && -- & -- & 21.0 & 7.4 \\ 
6 & -- & -- & 7e-08 & 0e+00 && -- & -- & 1381 & 699 && -- & -- & 23.3 & 12.2 \\ 
8 & -- & -- & 2e-08 & 0e+00 && -- & -- & 1856 & 747 && -- & -- & 38.2 & 15.4 \\ 
10 & -- & -- & 3e-08 & 0e+00 && -- & -- & 2352 & 1397 && -- & -- & 54.2 & 33.1 \\ 
12 & -- & -- & 4e-07 & 0e+00 && -- & -- & 3208 & 2442 && -- & -- & 77.7 & 66.6 
\\[-1pt] \bottomrule
 	\end{tabular}
 \end{table}
To evaluate the efficiency of the proposed ABAL method (i.e., Algorithm \ref{algo:abalmp}), we compare it with the state-of-the-art SeDuMi solver in the package CVX \cite{cvx,gb08}, similar to the approach in \cite{liu2022cramer}, the turning-free PDHG method \cite{wang2024tuning}, and the BAL method \eqref{alg:BALM-p} (with a constant $\gamma$). The  compared methods are denoted as ``ABAL'', ``SeDuMi'',  ``TF-PDHG'', and ``BAL-C'', respectively. Note that it is generally difficult to compare the solutions returned by different algorithms for solving constrained optimization problems, as both the violation of constraints and the objective value are important metrics. To make the comparison of different algorithms fair, we propose a way of obtaining a feasible solution to problem \eqref{eq:prob} from the points returned by different algorithms and focus on comparing their objective vaues. In particular, we propose to replace $\sigma_C^2$ in problem \eqref{eq:prob} with $(1+\epsilon)\sigma_C^2$ and solve the modified problem to satisfy a particular stopping criterion. Please see Appendix \ref{appen:feasible} for more details on this. In our simulations, we set $\epsilon = 10^{-3}$ and generate the parameters such as the channel vectors in problem \eqref{eq:prob} as in \cite{liu2022cramer}.

 The comparison results, averaged over $100$ Monte-Carlo runs for $N = 32, 64$ are reported in Table I. In Table I, “f-gap” denotes the relative objective value corresponding to the best one,  “iter” denotes the iteration number, and ``time'' is the running time in seconds. Additionally, the symbol ``--'' indicates that TF-PDHG fails to return a solution within 10,000 iterations or SeDuMi requires more than 2,000 seconds to terminate. From Table I, we can make the following observations: 
 (i) Our proposed ABAL outperforms the other two first-order methods, TF-PDHG and BAL-C. More specifically, ABAL is always faster than BAL-C, primarily due to its adaptive stepsize, while BAL-C uses a constant one. Compared to TF-PDHG which also utilizes an adaptive stepsize, the advantage of ABAL arises from its better way of updating the dual variable in \eqref{dualupdate}, as discussed in Section \ref{subsection:abal}. 
 (ii)  Our proposed ABAL is significantly faster than SeDuMi, achieving speedups ranging from 2.8 to 600 times. Notably, the performance advantage is more outstanding when $N = 64,$ which is our interested massive MIMO case. Overall, the simulation results demonstrate that our proposed ABAL is efficient in solving the ISAC beamforming design problem \eqref{eq:prob}.

\appendices

\section{Proof of Theorem \ref{thm:conv}} \label{section:equiv}
	We prove the theorem in two steps. First, we claim that all optimal solutions of  problem \eqref{eq:prob}  must satisfy $\sum_{k=1}^{K+1} \tr\left(\bW_k\right) = P_T$. We show this by contradiction. Assume that  there exists an optimal solution $\bW^*$ such that the strict inequality holds, i.e.,
	\begin{equation}
	\sum_{k=1}^{K+1} \tr\left(\bW^*_k\right) < P_T. \nonumber
	\end{equation}
	We construct a new point $\widehat \bW^* = \varrho \bW^*$ with $\varrho = P_T/\tr\left(\bW^*_k\right) > 1$. It is simple to show that $\widehat \bW^*$ is feasible to problem \eqref{eq:prob} with $\sum_{k=1}^{K+1} \tr\big(\widehat\bW^*_k\big) = P_T.$ 
	 However, we have 
	\[
	\tr\left(\left(\sum_{k=1}^{K+1} \widehat{\bW}^*_k\right)^{-1}\right) < \tr\left(\left(\sum_{k=1}^{K+1} \bW^*_k\right)^{-1}\right).
	\]
	This contradicts the optimality  of $\bW^*$. Therefore, any optimal solution of problem \eqref{eq:prob} must satisfy  $\sum_{k=1}^{K+1} \tr\left(\bW_k\right) = P_T$.

	Next, we show the existence of at least one optimal solution to problem \eqref{eq:prob} that satisfies all SINR inequality constraints with equality, i.e.,  
\begin{equation} \label{equ:appendix:equiv}
\rho_k \iprod{\bQ_k}{\bW_k}- \sum\nolimits_{i=1}^{K+1} \iprod{\bQ_k}{\bW_i} = \sigma_C^2,~~\forall\,k.
\end{equation}
Assume, for contradiction, that there exists an optimal solution $\bW^*$ to problem \eqref{eq:prob} for which at least one strict inequality holds. Without loss of generality, we consider the case where $k = 1$, which leads to  the following strict inequality: 
	\[  
\rho_1 \iprod{\bQ_1}{\bW_1^*}- \sum\nolimits_{i=1}^{K+1} \iprod{\bQ_1}{\bW_i^*} > \sigma_C^2.
	\]  
 This can be written as  
	\[  
\iprod{\bQ_1}{\rho_1 \bW_1^* - \sum\nolimits_{i=1}^{K+1} \bW_i^*} > \sigma_C^2. 
	\] 
 By choosing 
 \[
 \hat \varrho = \frac{\iprod{\bQ_1}{\sum\nolimits_{i=1}^{K+1} \bW_i^*}  + \sigma_C^2}{\iprod{\bQ_1}{\rho_1 \bW_1^*}} \in (0,1), 
 \]
 we construct the following new  point $\widehat \bW^*$:
 \[
 \begin{aligned}
 &\widehat{\bW}^*_1 := \hat \varrho \bW^*_1,~~ \widehat{\bW}^*_{K+1} :=  \bW^*_{K+1} + (1 - \hat \varrho)\bW^*_1, \\
 &\widehat{\bW}^*_k := \bW^*_k, ~~ k=2,3,\ldots,K.
 \end{aligned}
 \]
It is straightforward to verify that the above $\widehat{\bW}^*$ is feasible to problem \eqref{eq:prob} and that the first SINR inequality constraint of problem \eqref{eq:prob} is satisfied with equality. Moreover, we have 
 \[
 \tr\left(\left(\sum_{k=1}^{K+1} \widehat{\bW}^*_k\right)^{-1}\right)=\tr\left(\left(\sum_{k=1}^{K+1} \bW^*_k\right)^{-1}\right),
 \]
 which implies that the feasible solution $\widehat \bW^*$ is also optimal. By repeatedly applying the above procedure (at most $K$ times), we can construct a new optimal solution that satisfies all the equalities in \eqref{equ:appendix:equiv}.  
 The proof is complete.

\section{Customized Application of Algorithm \ref{algo:abalmp} to Problem \eqref{eq:prob2}} \label{appendix:customized}
In this part, we detail the customized application of Algorithm \ref{algo:abalmp} to problem  \eqref{eq:prob2} as discussed in Section \ref{section:customized}. Let us first introduce some notations used in this part. We use $\mathbf{0}_p$, $\mathbf{1}_p$, and $\bI_p$ to denote the $p$-dimensional all-zero vector,  the $p$-dimensional all-one vector, and the $p \times p$ identity matrix, respectively. The operation $\mvec(\cdot)$ converts a matrix into a column vector by stacking its columns and $\mat(\cdot)$ denotes the inverse operation of $\mvec(\cdot)$. 

Let $\bu=[\bx^\msH,\by^\msH]^\msH \in \mathbb{C}^{(K+2)N^2}$ with 
$$\bx = \begin{pmatrix} \mvec(\bW_1) \\ \vdots \\ \mvec(\bW_{K+1})\end{pmatrix} \in \mathbb{C}^{(K+1)N^2}~\text{and}~\by = \mvec(\bZ) \in \Cbb^{N^2}.$$ Then problem  \eqref{eq:prob2} can be equivalently reformulated as the form of problem \eqref{eq:gmodel} with the following $f$, $\bD$, and $\bb$: 
\setcounter{equation}{17}
\begin{figure*}[t]\normalsize
	\begin{equation}\label{inverse} \begin{aligned}	
	(\bA\bA^\msH + \bB\bB^\msH +  \theta^2 \bI_{K + n^2})^{-1}
	\bp^{t+1}
	={}&(\bA\bA^\msH + \bB\bB^\msH +  \theta^2 \bI_{K + n^2})^{-1}
	\begin{bmatrix}\br^{t+1} \\ \mvec(\bR^{t+1})\end{bmatrix} \\[10pt]
	={}&
	\begin{bmatrix}
	\bL^{-1} \left(\br^{t+1} - \frac{\bT_{12} \mvec(\bR^{t+1})}{K + 2 + \theta^2}  \right) \\
	\frac{1}{K+2+\theta^2} \left(\mvec(\bR^{t+1})  - \bT_{12}^\msH \bL^{-1} \Big(\br^{t+1} - \frac{\bT_{12} \mvec(\bR^{t+1})}{K + 2 + \theta^2}  \Big)\right)
	\end{bmatrix}.
	\end{aligned}
	\end{equation}
	\hrulefill
\end{figure*}
 \setcounter{equation}{13}
\begin{itemize}
	\item The function $f(\bu)=f_1(\bx)+f_2(\by)$ with  
 $f_1(\bx) = \mathbb{I}_{\Xcal}(\bx)$, $f_2(\by) = \tr(\mat(\y)^{-1}) + \mathbb{I}_{\Ycal}(\by)$ where 
	$$\begin{aligned}
	&\Xcal:= \big\{\bx\in\mathbb{C}^{(K+1)N^2}\mid[\mat(\bx_{1:N^2}), \mat(\bx_{N^2 + 1: 2N^2}), \nonumber\\ 
	&\qquad\qquad \qquad\qquad\ldots,\mat(\bx_{K N^2 +1: (K+1)N^2})]^\msH \in \Wcal \big\},\nonumber\\
	&\Ycal:= \big\{\by \in \Cbb^{N^2} \mid \mat(\by) \in \mathbb{H}_{+}^{N \times N}\big\}.
	\end{aligned}$$   
 Here, the notation $\bx_{i_1:i_2}$ denotes a subset of the vector $\bx$, consisting of all elements from the $i_1$-th element to the $i_2$-th element.
	\item The coefficient matrix $\bD=[\bA,\bB]$ with 
	\begin{equation}\label{eq:blockA}
	\bA = \begin{pmatrix} 
	\bA_1 \\
	\bA_2
	\end{pmatrix} 
	\in \Cbb^{(K + N^2)\times (K+1)N^2},
	\end{equation}
	where 
	\begin{equation} 
	\setlength{\arraycolsep}{0.5pt}
	\begin{aligned}
	&\bA_1 = 
	\begin{pmatrix}
	\rho_1 \mvec(\bQ_1)^\msH & \mathbf{0} & \cdots  & \mathbf{0} &  \mathbf{0}  \\
	\mathbf{0} & \rho_2 \mvec(\bQ_2)^\msH  & \cdots  & \mathbf{0} & \mathbf{0} \\
	& & \vdots & & \\ 
	\mathbf{0} & \mathbf{0} & \cdots  & \rho_K \mvec(\bQ_K)^\msH  & \mathbf{0} 
	\end{pmatrix}\nonumber\\
	&\hspace{16em}\in \Cbb^{K\times (K+1)N^2},
	\end{aligned} 
	\end{equation}
	\[
	\bA_2 = 
 \mathbf{1}_{K+1} \otimes \bI_{N^2}
	\in \Cbb^{N^2\times (K+1)N^2},
	\]
	and  
	\begin{equation} \label{eq:blockB}
	\bB =  
	\begin{pmatrix} 
	\bB_1 \\
	\bB_2
	\end{pmatrix} 
	\in \Cbb^{(K + N^2)\times  N^2},  
	\end{equation}
	with 
	\[ 
	\bB_1 = 
	-
	\begin{pmatrix}
	\mvec(\bQ_1)^\msH   \\
	\mvec(\bQ_2)^\msH   \\
	\vdots \\
	\mvec(\bQ_K)^\msH 
	\end{pmatrix}
	\in \Cbb^{K\times N^2},
	\bB_2 = 
	-\bI_{N^2}
	\in \Cbb^{N^2 \times N^2}. 
	\]
	\item The vector 
 \[
\bb =  \begin{pmatrix}\sigma_C^2 \mathbf{1}_K\\ \mathbf{0}_{N^2} \end{pmatrix}. 
\]
\end{itemize}

The proposed ABAL algorithm for solving problem \eqref{eq:prob2} is thus given as 
\begin{equation}\label{equ:update:xy:}
	\begin{cases}
 \widetilde\bx^t = \bx^t-\tau_{t-1}\bA^\msH \blam^t, \\
	\bx^{t+1}=\operatorname{prox}_{\tau_{t-1}f}\left(\widetilde \bx^t\right), ~ \Delta \bx^t = \bx^{t+1} - \widetilde \bx^t,\\
	\widetilde \by^t = \by^t-\tau_{t-1} \bB^\msH \blam^t, \\\by^{t+1}=\operatorname{prox}_{\tau_{t-1} g}\left(\widetilde \by^t\right), ~  \Delta \by^t = \by^{t+1} - \widetilde \by^t,\\
\eta_t = \proj_{[\underline{\eta}, \overline{\eta}]}\left({\frac{\sqrt{\|\bx^{t+1}\|_2^2 + \|\by^{t+1}\|_2^2}}{\sqrt{\|\Delta \bx^t\|_2^2+ \|\Delta \by^t\|_2^2+\theta^2\tau_{t-1}^2 \|\blam^t\|_\Fsf^2}}}\right), \\ 	
	 \kappa_t = 1-\omega_t+\omega_t \eta_t,~ \tau_t = \kappa_t \tau_{t-1}, \\
	 \bp^{t+1} =(1+ \kappa_t)(\bA \bx^{t+1} + \bB \bx^{t+1}) -  \kappa_t(\bA \bx^{t} + \bB \bx^{t}) -\bb,\\
	\blam^{t+1} =\blam^t+\frac{1}{\tau_t}\left(\bA \bA^\msH+\bB \bB^\msH+\theta^2 \bI_{K+N^2}\right)^{-1} \bp^{t+1}. 
	\end{cases}
 \end{equation}
In the above, the Lagrange multiplier  $\blam$ corresponds to $\bmu$ and $\bLam$ in \eqref{equ:customized}, while the variable $\bp$ corresonds to $\br$ and $\bR$ in \eqref{equ:customized} in the following manner: 
\begin{equation}
\label{equ:transform}
\blam = \begin{pmatrix}
\bmu \\ 
\mbox{vec}(\bLam)
\end{pmatrix},
~
\bp = \begin{pmatrix}
\br \\ 
\mbox{vec}(\bR)
\end{pmatrix}.
\end{equation}
The updates in the first eight lines in  
\eqref{equ:customized} can be easily derived.

Next we focus on the updates in the last two lines in  
\eqref{equ:customized}, i.e., we need to investigate the $\blam$-subproblem in \eqref{equ:update:xy:} in detail. At first glance, solving the $\blam$-subproblem in \eqref{equ:update:xy:} seems to require solving a  linear equation with the coefficient matrix $$\bA\bA^\msH + \bB\bB^\msH + \theta^2 \bI_{K + N^2} \in \mathbb{H}^{K+N^2},$$ which typically incurs a complexity of $\mathcal{O}((K+N^2)^3)$. However, by leveraging the block structure of $\bA$  in \eqref{eq:blockA} and $\bB$ in \eqref{eq:blockB}, we can significantly reduce this complexity to $\mathcal{O}(K^3 + KN^2)$. We provide the details below.

Through straightforward calculations, we have
\begin{align*}
&\bA\bA^\msH + \bB\bB^\msH + \theta^2 \bI_{K + N^2}\nonumber \\
&=\begin{pmatrix}
\bA_1 \bA_1^\msH + \left| \bH^\msH \bH \right|^2 + \theta^2 \bI_K & \bT_{12} \\
\bT_{12}^\msH & (K+2+ \theta^2) \bI_{N^2}
\end{pmatrix},\nonumber 
\end{align*}
where
\[
\bA_1 \bA_1^\msH = \Diag(\rho_1^2 \|\bh_1\|^4, \ldots, \rho_K^2 \|\bh_K\|^4)
\]
and 
\[
\bT_{12} = \begin{pmatrix}
(\rho_1 + 1) \mvec(\bQ_1)^\msH \\
\vdots \\
(\rho_K + 1) \mvec(\bQ_K)^\msH
\end{pmatrix}.
\]
Let \begin{align*}
\bL&=\bA_1 \bA_1^\msH + \left| \bH^\msH \bH \right|^2 + \theta^2 \bI_K - \frac{1}{K + 2 + \theta^2} \bT_{12} \bT_{12}^\msH \nonumber\\
&=\theta^2 \bI_K + \left| \bH^\msH \bH \right|^2 \odot (\Diag(\brho \odot \brho) + \bS),
\end{align*}
where $\bH = [\bh_1, \bh_2, \ldots, \bh_K]$ and
\[
\bS = \1_K\1_K^\msH - (\brho + \1_K) (\brho + \1_K)^\msH/(K + 2 + \theta^2). 
\]
Therefore, we can express the inverse of $\bA\bA^\msH  +  \bB\bB^\msH  + \theta^2 \bI_{K + n^2}$ as
\begin{align*}
&\left(\bA\bA^\msH  +  \bB\bB^\msH  + \theta^2 \bI_{K + n^2}\right)^{-1} \nonumber \\[2pt]
={}&\begin{bmatrix}
\bL^{-1}  & -\frac{\bL^{-1}  \bT_{12}}{K + 2 + \theta^2}  \\[3pt]
-\frac{\bT_{12}^\msH \bL^{-1} }{K + 2 + \theta^2} 
&  \frac{\bT_{12}^\msH \bL^{-1} \bT_{12}}{(K + 2 + \theta^2)^2}  + 
\frac{\bI_{N^2}}{K + 2 + \theta^2} 
\end{bmatrix}.
\nonumber 
\end{align*} Using the transformation of $\bp$ in \eqref{equ:transform}, we have \eqref{inverse} on top of this page. Consequently, to solve the $\blam$-subproblem, we only need solve the following linear equation:  
$$
\bL \mathbf{q} = \br^{t+1} - \frac{1}{K + 2 + \theta^2} \bT_{12} \mvec(\bR^{t+1}),
$$
whose complexity is $\mathcal{O}(K^3 + KN^2)$.

Finally, using the transformation of $\blam$ in 
\eqref{equ:transform}, we can derive the last two updates in \eqref{equ:customized} through some straightforward calculations.

\setcounter{equation}{18}
\section{Feasibility of Returned Solutions by Different Algorithms}\label{appen:feasible}
Let $\bW$ be an approximate solution to problem \eqref{eq:prob2} with $\sigma_C^2$ replaced by $(1+\epsilon)\sigma_C^2$, returned by a specific algorithm, satisfying the following criterion:
\begin{equation}\label{equ:sectionc:a0}
\max\left\{\big\|\mathcal{A}(\bW, \bZ) - (1 + \epsilon) \sigma_C^2 \mathbf{1}_K\big\|_{\Fsf}, \big\|\mathcal{B}(\bW,\bZ)\big\|_{\Fsf}\right\}
\leq \mathrm{tol},
\end{equation}
where 
\begin{equation}\label{tol}
\mathrm{tol}:= \frac{\epsilon \sigma_C^2}{1 + \min_k \|\bh_k\|^2}.
\end{equation}
Using  \eqref{equ:sectionc:a0}, $\bQ_k = \bh_k \bh_k^\msH$, and the definitions of $\mathcal{A}$ and $\mathcal{B}$ given after \eqref{eq:prob2}, for any $k \in \{1, 2, \ldots, K\}$, we have 
\begin{align}
{}& \rho_k \iprod{\bQ_k}{\bW_k}- \sum\nolimits_{i=1}^{K+1} \iprod{\bQ_k}{\bW_i} - \sigma_C^2 \nonumber\\
={}&(\mathcal{A}(\bW, \bZ))_k - (1 + \epsilon) \sigma_C^2 + \epsilon \sigma_C^2  
  +  \bh_k^\msH\left(\Sigma_{k = 1}^{K+1} \bW_k - \bZ\right) \bh_k \nonumber \\
\geq{}& - \mathrm{tol} + \epsilon \sigma_C^2 - \|\bh_k\|^2 \mathrm{tol} \nonumber \\
={}& \epsilon \sigma_C^2 - ( 1 + \|\bh_k\|^2) \mathrm{tol} \nonumber \\
\geq{}&0,  \nonumber
\end{align} where the first inequality is due to \eqref{equ:sectionc:a0} and the second inequality is due to the choice of $\mathrm{tol}$ in \eqref{tol}. This shows that $\bW$ in \eqref{equ:sectionc:a0} is feasible to problem \eqref{eq:prob}. 
\ifCLASSOPTIONcaptionsoff
\newpage
\fi
\newpage

\bibliographystyle{IEEEtran}
\bibliography{ref_liu_v4,IEEEabrv}
\begin{thebibliography}{10}
\providecommand{\url}[1]{#1}
\csname url@samestyle\endcsname
\providecommand{\newblock}{\relax}
\providecommand{\bibinfo}[2]{#2}
\providecommand{\BIBentrySTDinterwordspacing}{\spaceskip=0pt\relax}
\providecommand{\BIBentryALTinterwordstretchfactor}{4}
\providecommand{\BIBentryALTinterwordspacing}{\spaceskip=\fontdimen2\font plus
\BIBentryALTinterwordstretchfactor\fontdimen3\font minus
  \fontdimen4\font\relax}
\providecommand{\BIBforeignlanguage}[2]{{%
\expandafter\ifx\csname l@#1\endcsname\relax
\typeout{** WARNING: IEEEtran.bst: No hyphenation pattern has been}%
\typeout{** loaded for the language `#1'. Using the pattern for}%
\typeout{** the default language instead.}%
\else
\language=\csname l@#1\endcsname
\fi
#2}}
\providecommand{\BIBdecl}{\relax}
\BIBdecl

\bibitem{Tomioka2009}
R.~Tomioka and M.~Sugiyama, ``Dual-augmented {L}agrangian method for efficient
  sparse reconstruction,'' \emph{IEEE Signal Process. Lett.}, vol.~16, no.~12,
  pp. 1067--1070, 2009.

\bibitem{aybat2012first}
N.~S. Aybat and G.~Iyengar, ``A first-order augmented {L}agrangian method for
  compressed sensing,'' \emph{SIAM J. Optim.}, vol.~22, no.~2, pp. 429--459,
  2012.

\bibitem{Mota2012}
J.~F.~C. Mota, J.~M.~F. Xavier, P.~M.~Q. Aguiar, and M.~Puschel, ``Distributed
  basis pursuit,'' \emph{IEEE Trans. Signal Process.}, vol.~60, no.~4, pp.
  1942--1956, 2012.

\bibitem{yang2013linearized}
J.~Yang and X.~Yuan, ``Linearized augmented {L}agrangian and alternating
  direction methods for nuclear norm minimization,'' \emph{Math. Comp.},
  vol.~82, no. 281, pp. 301--329, 2013.

\bibitem{Chang2016}
T.-H. Chang, M.~Hong, W.-C. Liao, and X.~Wang, ``Asynchronous distributed
  {ADMM} for large-scale optimization—part {I}: Algorithm and convergence
  analysis,'' \emph{IEEE Trans. Signal Process.}, vol.~64, no.~12, pp.
  3118--3130, 2016.

\bibitem{liu2022cramer}
F.~Liu, Y.-F. Liu, A.~Li, C.~Masouros, and Y.~C. Eldar, ``Cram\'er-{R}ao bound
  optimization for joint radar-communication beamforming,'' \emph{IEEE Trans.
  Signal Process.}, vol.~70, pp. 240--253, 2022.

\bibitem{liu2024survey}
\BIBentryALTinterwordspacing
Y.-F. Liu, T.-H. Chang, M.~Hong, Z.~Wu, A.~M.-C. So, E.~A. Jorswieck, and
  W.~Yu, ``A survey of recent advances in optimization methods for wireless
  communications,'' \emph{IEEE J. Sel. Areas Commun.}, 2024. [Online].
  Available: \url{http://dx.doi.org/10.1109/JSAC.2024.3443759}
\BIBentrySTDinterwordspacing

\bibitem{douglas1956numerical}
J.~Douglas and H.~H. Rachford, ``On the numerical solution of heat conduction
  problems in two and three space variables,'' \emph{Trans. Am. Math. Soc.},
  vol.~82, no.~2, pp. 421--439, 1956.

\bibitem{hestenes1969multiplier}
M.~R. Hestenes, ``Multiplier and gradient methods,'' \emph{J. Optim. Theory
  Appl.}, vol.~4, no.~5, pp. 303--320, 1969.

\bibitem{powell1969method}
M.~J. Powell, ``A method for nonlinear constraints in minimization problems,''
  \emph{Optim.}, pp. 283--298, 1969.

\bibitem{chambolle2011first}
A.~Chambolle and T.~Pock, ``A first-order primal-dual algorithm for convex
  problems with applications to imaging,'' \emph{J. Math. Imaging Vis.},
  vol.~40, pp. 120--145, 2011.

\bibitem{he2021balanced}
B.~He and X.~Yuan, ``Balanced augmented {L}agrangian method for convex
  programming,'' \emph{arXiv:2108.08554}, 2021.

\bibitem{liu2021acceleration}
Y.~Liu, Y.~Xu, and W.~Yin, ``Acceleration of primal--dual methods by
  preconditioning and simple subproblem procedures,'' \emph{J. Sci. Comput.},
  vol.~86, no.~2, p.~21, 2021.

\bibitem{Ma2023}
\BIBentryALTinterwordspacing
Y.~Ma, X.~Cai, B.~Jiang, and D.~Han, ``Understanding the convergence of the
  preconditioned {PDHG} method: A view of indefinite proximal {ADMM},''
  \emph{J. Sci. Comput.}, vol.~94, no.~3, 2023. [Online]. Available:
  \url{http://dx.doi.org/10.1007/s10915-023-02105-9}
\BIBentrySTDinterwordspacing

\bibitem{monteiro2024lowrank}
R.~D.~C. Monteiro, A.~Sujanani, and D.~Cifuentes, ``A low-rank augmented
  {L}agrangian method for large-scale semidefinite programming based on a
  hybrid convex-nonconvex approach,'' \emph{arXiv:2401.12490}, 2024.

\bibitem{LiuLMOR}
Y.-F. Liu, X.~Liu, and S.~Ma, ``On the nonergodic convergence rate of an
  inexact augmented {L}agrangian framework for composite convex programming,''
  \emph{Math. Oper. Res.}, vol.~44, no.~2, pp. 632--650, 2019.

\bibitem{Lorenz2019}
D.~A. Lorenz and Q.~Tran-Dinh, ``Non-stationary {D}ouglas--{R}achford and
  alternating direction method of multipliers: Adaptive step-sizes and
  convergence,'' \emph{Comput. Optim. Appl.}, vol.~74, no.~1, pp. 67--92, 2019.

\bibitem{Zhang2024}
\BIBentryALTinterwordspacing
T.~Zhang, Y.~Xia, and S.~Li, ``${O}(1/k^2)$ convergence rates of (dual-primal)
  balanced augmented {L}agrangian methods for linearly constrained convex
  programming,'' \emph{Numer. Algorithms}, 2024. [Online]. Available:
  \url{http://dx.doi.org/10.1007/s11075-024-01796-x}
\BIBentrySTDinterwordspacing

\bibitem{ITU-R2023}
\emph{Framework and Overall Objectives of the Future Development of {IMT} for
  2030 and Beyond}, ITU-R Std. M.2160-0, Nov. 2023.

\bibitem{liu2022integrated}
F.~Liu, Y.~Cui, C.~Masouros, J.~Xu, T.~X. Han, Y.~C. Eldar, and S.~Buzzi,
  ``Integrated sensing and communications: Toward dual-functional wireless
  networks for 6{G} and beyond,'' \emph{IEEE J. Sel. Areas Commun.}, vol.~40,
  no.~6, pp. 1728--1767, 2022.

\bibitem{liu2018toward}
F.~Liu, L.~Zhou, C.~Masouros, A.~Li, W.~Luo, and A.~Petropulu, ``Toward
  dual-functional radar-communication systems: Optimal waveform design,''
  \emph{IEEE Trans. Signal Process.}, vol.~66, no.~16, pp. 4264--4279, 2018.

\bibitem{vijay2019toward}
K.~V. Mishra, M.~Bhavani~Shankar, V.~Koivunen, B.~Ottersten, and S.~A.
  Vorobyov, ``Toward millimeter-wave joint radar communications: A signal
  processing perspective,'' \emph{IEEE Signal Process. Mag.}, vol.~36, no.~5,
  pp. 100--114, 2019.

\bibitem{liu2020joint}
X.~Liu, T.~Huang, N.~Shlezinger, Y.~Liu, J.~Zhou, and Y.~C. Eldar, ``Joint
  transmit beamforming for multiuser {MIMO} communications and {MIMO} radar,''
  \emph{IEEE Trans. Signal Process.}, vol.~68, pp. 3929--3944, 2020.

\bibitem{liu2024random}
S.~Lu, F.~Liu, F.~Dong, Y.~Xiong, J.~Xu, Y.-F. Liu, and S.~Jin, ``Random {ISAC}
  signals deserve dedicated precoding,'' \emph{IEEE Trans. Signal Process.},
  vol.~72, pp. 3453--3469, 2024.

\bibitem{wu2024efficient}
J.~Wu, Z.~Wang, Y.-F. Liu, and F.~Liu, ``Efficient global algorithms for
  transmit beamforming design in {ISAC} systems,'' \emph{IEEE Trans. Signal
  Process.}, 2024.

\bibitem{zou2024energy}
J.~Zou, S.~Sun, C.~Masouros, Y.~Cui, Y.-F. Liu, and D.~W.~K. Ng,
  ``Energy-efficient beamforming design for integrated sensing and
  communications systems,'' \emph{IEEE Trans. Commun.}, vol.~72, no.~6, pp.
  3766--3782, 2024.

\bibitem{o2020equivalence}
D.~O’Connor and L.~Vandenberghe, ``On the equivalence of the primal-dual
  hybrid gradient method and {D}ouglas--{R}achford splitting,'' \emph{Math.
  Program.}, vol. 179, no.~1, pp. 85--108, 2020.

\bibitem{wang2024tuning}
Y.~Wang, H.~Lan, and Y.~Ye, ``A tuning-free primal-dual splitting algorithm for
  large-scale semidefinite programming,'' \emph{arXiv:2402.00311}, 2024.

\bibitem{cvx}
M.~Grant and S.~Boyd, ``{CVX}: Matlab software for disciplined convex
  programming, version 2.1,'' \url{https://cvxr.com/cvx}, 2014.

\bibitem{gb08}
------, ``Graph implementations for nonsmooth convex programs,'' in
  \emph{Recent Advances in Learning and Control}, ser. Lecture Notes in Control
  and Information Sciences, V.~Blondel, S.~Boyd, and H.~Kimura, Eds.\hskip 1em
  plus 0.5em minus 0.4em\relax Springer-Verlag Limited, 2008, pp. 95--110,
  \url{http://stanford.edu/~boyd/graph_dcp.html}.

\end{thebibliography}
\end{document}